# Charting a Path to Efficient Onboarding: The Role of Software Visualization

Fernando Padoan
CESAR School
Recife, PE, Brazil
fop@cesar.school

Ronnie de Souza Santos
University of Calgary
Calgary, AB, Canada
ronnie.desouzasantos@ucalgary.com

Rodrigo Pessoa Medeiros
Instituto Federal da Paraíba
Cabedelo, PB, Brazil
rodrigo.medeiros@ifpb.edu.br

## ABSTRACT

**Background**. Within the software industry, it is commonly estimated that software professionals invest a substantial portion of their work hours in the process of understanding existing systems. In this context, an ineffective technical onboarding process, which introduces newcomers to software under development, can result in a prolonged period for them to absorb the necessary knowledge required to become productive in their roles. **Goal**. The present study aims to explore the familiarity of managers, leaders, and developers with software visualization tools and how these tools are employed to facilitate the technical onboarding of new team members. **Method**. To address the research problem, we built upon the insights gained through the literature and embraced a sequential exploratory approach. This approach incorporated quantitative and qualitative analyses of data collected from practitioners using questionnaires and semi-structured interviews. **Findings**. Our findings demonstrate a gap between the concept of software visualization and the practical use of onboarding tools and techniques. Overall, practitioners do not systematically incorporate software visualization tools into their technical onboarding processes due to a lack of conceptual understanding and awareness of their potential benefits. **Conclusion**. The software industry could benefit from standardized and evolving onboarding models, improved by incorporating software visualization techniques and tools to support program comprehension of newcomers in the software projects.

## KEYWORDS

software visualization, technical onboarding, program comprehension

## 1 INTRODUCTION

In the software industry, technical onboarding is the process of integrating newly hired or transitioning software professionals into software development teams [2, 5]. This process is focused on equipping these individuals with the essential skills, knowledge, tools, and workflows used in software projects. For software engineers, including developers and programmers, a successful technical onboarding experience is crucial to support them in program comprehension and enable them to swiftly become productive and collaborative team members, thereby enhancing their effectiveness in contributing to software development projects [30, 31].

Comprehending code and the associated software system requirements are pivotal aspects of software development. It is worth noting that developers dedicate a significant portion of their time to reviewing and troubleshooting existing code [39]. In this context, various factors, including outdated documentation, overly complex code, a growing number of features, tight project timelines, and alterations in team composition, can prolong the duration required for new team members to achieve their maximum potential. This challenging situation, in turn, may prompt them to leave their teams due to low satisfaction [33].

Recent research highlights the knowledge management challenges faced by software engineering organizations. These challenges are rooted in the necessity to maintain access to knowledge originating from the conceptual and construction phases of software development, even in the face of team adjustments, such as when team members depart from the project. In other words, it is essential to preserve and manage the knowledge generated during their participation in the project [4, 41].

In general, the processes of technical onboarding and knowledge management can benefit significantly from the use of appropriate tools [12, 34]. This includes the integration of software visualization tools [34], which encompass various methods and techniques for representing and conveying information about algorithms, software programs, and project data [17]. Such tools are particularly effective in enhancing program comprehension during the technical onboarding phase within development teams [34, 40].

In this study, we have two primary objectives: 1) To evaluate the familiarity of software project managers, technical leaders, and software developers with the concept of software visualization, and 2) To investigate the practical implementation of tools incorporating these concepts during the technical onboarding of new team members. In this context, we aim to address the following overarching research question: ***RQ***. *How do software professionals navigate technical onboarding and the utilization of software visualization tools within the software industry?* To narrow our research focus, we have formulated the following specific inquiries:

> ***RQ1***. *How familiar are software professionals with software visualization tools?*
>
> ***RQ2***. *How frequently do software companies incorporate software visualization tools into their technical onboarding processes?*
>
> ***RQ3***. *How do software professionals assess the effectiveness of software visualization tools in improving their comprehension of source code during the onboarding process?*

From this introduction, our study is organized as follows. Section 2 explores concepts around software visualization tools. In Section 3, we describe the details of our research method. In Section 4, we present our findings. In Section 5, we present the implications of our study. Lastly, Section 6 summarizes our contributions.

## 2 BACKGROUND

In this section, we introduce two core concepts central to our study, namely, onboarding and software visualization. We start by outlining overarching definitions of the general onboarding process, then proceed by presenting specific aspects related to onboarding in software engineering. We finish highlighting concepts around software visualization and program comprehension and its relation to onboarding.

### 2.1 Onboarding New Employees

Onboarding, the process by which new hires are introduced as members of a team or organization, may be applied according to formal, well-defined methodologies, such as the research-based model proposed by Bauer [2] to support the design of institutionalized onboarding programs, from the observation that such programs, with formal tactics, are more successful than individualized onboarding alone [5]. Bauer's model also takes into consideration four distinct levels ("Four C's") of onboarding, namely: *Compliance* — learning basic legal and policy-related rules and regulations; *Clarification* — understand one's job and its expectations; *Culture* — being provided a sense of organizational norms, formal and informal; and *Connection* — establishing interpersonal relationships and information networks [2].

Research indicates that effective organizational socialization plays an important role in turning newcomers into engaged employees. This process is influenced by several factors, where new employees are active participants, seeking information and feedback in order to socially adjust [3]. Organizations also influence these factors by implementing orientation programs and pairing newcomers with mentors — insiders that support the newly hired employee by helping them navigate the organization, offering advice, or answering uncomfortable questions they may have [3, 38]. Other studies indicate that, although with institutionalized onboarding tactics, the learning and adjustment of new employees result in higher levels of fit and lower turnover, the use of individualized tactics relates to higher innovation and to demonstrating employees' personal values [3].

The aspect of self-efficacy, or how confident new employees feel about performing their job well, is considered to have a major impact on their commitment to the organization and on job satisfaction [2]. Onboarding in complex settings or systems also demands time from current and previous members of a team and when inefficient, these processes can lead to high turnover in an organization [3, 6]. Therefore, the productivity of developers and their teams can be seen as both an individual and managerial challenge [15].

### 2.2 Onboarding in Software Engineering

In software engineering organizations, onboarding is usually twofold, where a first non-technical session for all roles, aimed at training new employees in the various policies and cultural aspects of the organization, is followed by a technical onboarding for developers, with activities that expose them to details of software architecture and coding standards adopted in the construction of the solutions in which they must work [31, 38].

One challenge faced by software engineering organizations among others, is knowledge management, or how to ensure that the knowledge generated during the design and construction of systems is possible to recover and act on, even after months of development and changes in the team, as well as the need to consider not only explicit knowledge, but also tacit knowledge [4, 41].

Another challenge these organizations face in global software development and hybrid work contracts, and thus aggravated by the recent COVID-19 pandemic, is the socialization of developers hired under remote work arrangements or during lockdown restrictions [15, 30]. Developers often report the lack of systematized onboarding processes, lack of opportunities to adjust to company culture, the need for longer mentoring periods, or even dissatisfaction with the behavior of colleagues towards remote conference etiquette, which may hinder collaboration and affect remote work experience negatively [5, 30, 31].

### 2.3 Software Visualization and Program Comprehension

Having its origins in Information Visualization, the sub-field of Software Visualization is a domain concerning visual representations of various properties of software systems and their development process [10]. Several taxonomies of software visualization research and tools have been proposed, such as Myers' matrix, that distinguishes static and dynamic visualization types, then classifies visualizations in levels of abstraction from data (lowest) to code (higher) and algorithms (highest level) [27], as shown in Figure 1.

|  | Static | Dynamic |
|---|---|---|
| Data |  |  |
| Code |  |  |
| Algorithm |  |  |

**Figure 1: Myers' Taxonomy of Software Visualization**

From its beginning until the end of the nineties, its primary focus was in aiding the comprehension of programs and their behavior, having subsequently evolved to capture and represent the evolution not only of source code and data but also requirements and developer contributions [11, 17, 29]. Examples of modern visualization metaphors include *software cities* for visualizing the evolution of code structure [32], *hierarchical edge bundles* that represent dependencies among code units [18], or activities that flow from module to module in colorful *developer rivers* [7], to name only a few.

Recently, new approaches have emerged to facilitate and support program comprehension activities. These approaches introduce concepts like 'Learnable Programming' and 'Moldable Development,' which provide dynamic ways to interact with code rather than solely depending on traditional text editors and programming IDEs [8, 37]. Furthermore, the integration of modern augmented and virtual reality devices into software visualization has opened up new possibilities for tools and solutions that work at the intersection of these domains [23]. Since these visualization tools are designed to represent aspects such as system structure, functionality behavior, and software evolution, they have the potential to not only support program comprehension but also help newcomers achieve satisfactory levels of productivity [10].

Despite the potential of software visualization tools to improve program comprehension through visual techniques, they remain underutilized in the software industry [36]. This underutilization can be attributed in part to the limited awareness of commercial tools of this nature within software development organizations [35]. Recent research in software visualization has primarily focused on academic contexts or used these tools as proof of concept in open-source systems, with limited available materials reporting their practical use in the industry [21, 22].

## 2.4 Research Opportunity

In a mapping study analyzing 31 unique papers published within the last five years [citation hidden for blind review], the analysis of the state-of-the-art research in software visualization related to technical onboarding reveals that studies primarily focus on tool proposals. These findings highlighted a notable gap in studies that concentrate on the evaluation and analysis of software visualization in real-world settings. This underscores the need for research in the industrial context and points to opportunities for exploring the specific needs of software professionals and the effectiveness of software visualization tools in technical onboarding processes.

## 3 METHOD

The scarcity of comprehensive research in current literature regarding the use of software visualization tools and techniques in technical onboarding within software teams conducted in the industrial context has led us to employ a mixed-method approach [13, 14]. This approach was designed to collect and analyze a substantial body of evidence emerging from real-world industry practices. To accomplish this, we applied a quantitative methodology using a survey [25] to collect data from a sample of industry practitioners via a questionnaire. Additionally, we have gathered qualitative sources from the experiences of software professionals by conducting scripted interviews [19].

Mixed methods approaches offer practical advantages when dealing with complex research questions, providing valuable insights into emerging and unexpected topics [13]. Our choice to collect quantitative data before qualitative data, following a sequential exploratory strategy, enabled us to utilize the qualitative data to enhance the interpretation of the quantitative findings. Figure 2 illustrates our research design.

## 3.1 Survey

The survey method has a rich history in social research, spanning decades and extending its applicability to numerous academic domains, including healthcare, politics, psychology, and sociology. In software engineering, they are widely used to explore the insights, practices, experiences, and opinions of software practitioners, providing a systematic and efficient way to collect valuable data from this population [20, 26]. In the first phase of this study, we used well-established guidelines to conduct surveys in software engineering [20] and assessed the perspectives and opinions of software professionals about software visualization and technical onboarding.

### 3.1.1 Questionnaire.

The questionnaire was constructed based on insights derived from evidence collected in previous research on software visualization and technical onboarding. It consisted of a total of 18 multiple-choice questions, e.g., close-ended questions. Since our subsequent phase was entirely qualitative, open-ended questions were not included in the questionnaire.

The questions were organized into three main sections. The first section, *About You and Your Workplace*, aimed to collect demographic information about the participants. The second section, *Onboarding*, focused on their understanding of technical onboarding and their perspectives on this practice. The third section, *Software Visualization*, was tailored to assess the the participants' familiarity with program comprehension and the tools that support it. We concluded the questionnaire by informing participants that our study would have a subsequent stage involving interviews. We invited those who were interested in participating in the interview phase to provide their contact email for further communication. The final questionnaire is presented in the Table 1.

The phrasing of the survey questions and the options for responses was crafted based on both the evidence gathered from the literature and the experience of one of the researchers who belongs to the participant population, namely, software professionals. Through a pilot validation, the questionnaire was deemed suitable for the target audience.

### 3.1.2 Sampling and Data Collection.

In our data collection process, we employed two sampling methods following the guidance of [1]: convenience sampling and snowballing. Convenience sampling involves selecting participants who are readily available and willing to participate in the research. On the other hand, purposive sampling entails selecting participants with specific characteristics relevant to the study who often belong to a specific site or unit.

For convenience sampling, we identified potential participants within the software industry who showed an interest in participating in our study. Leveraging our extensive network of software professionals, we distributed the questionnaire to them through various channels, including email, chat, and social media.

Additionally, for purposive sampling, we used the communication channels of a prominent software company in South America to promote our questionnaire. This company, established in 1996, specializes in on-demand software development and serves clients across diverse sectors, including finance, telecommunications, manufacturing, and services, and is known for its onboarding processes.

Data collection started in early March 2023 and continued for the following three weeks. Throughout this timeframe, we consistently promoted our research to gather as many participants as possible. By the end of this data collection period, we had received responses from 41 participants. This sample featured individuals from various companies, representing diverse backgrounds and occupying distinct technical roles within software development teams.

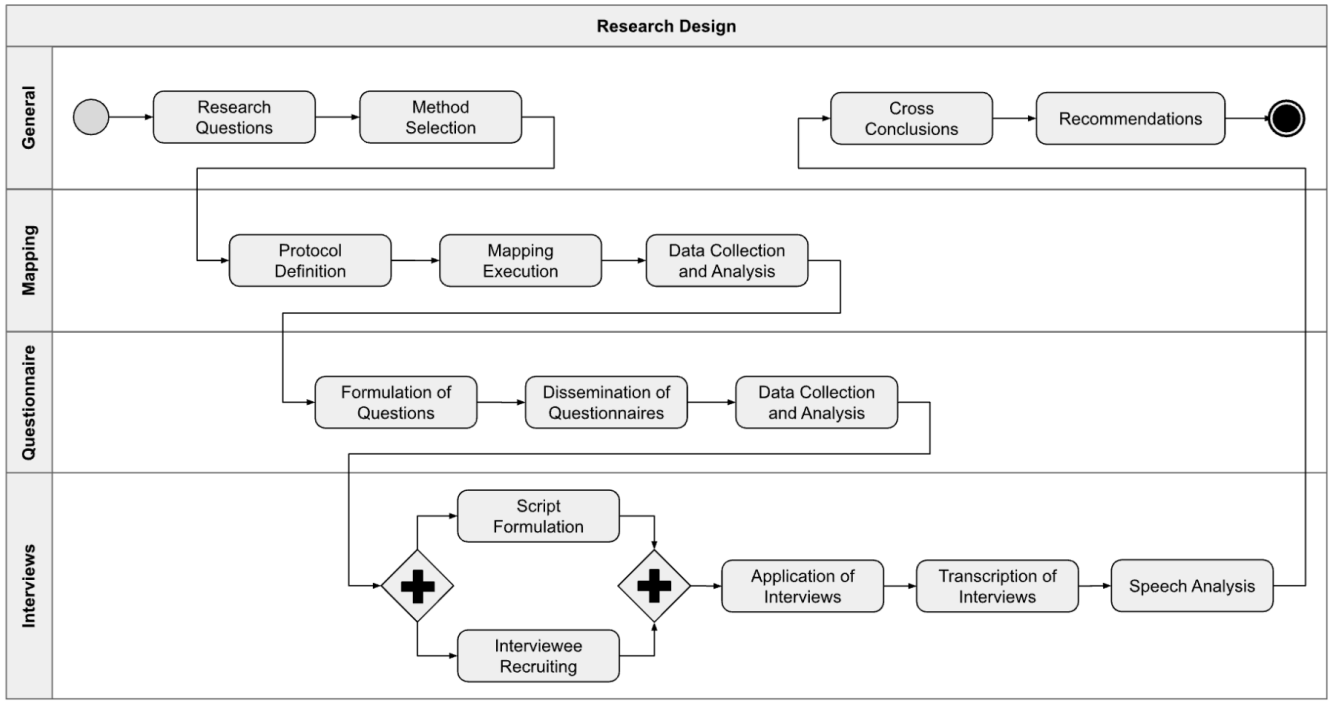

Figure 2: Mixed-Method Research Design

*3.1.3 Data Analysis.*
After collecting quantitative data, we applied the principles of descriptive statistics, following the guidelines outlined in [16]. This approach enabled us to systematically explore and summarize the distribution and frequency of responses gathered from the questionnaires. Descriptive statistics provided a comprehensive and insightful view of the data, aiding in the understanding of key aspects and trends within the sample. This quantitative analysis laid the foundation for a more effective qualitative stage of our research.

### 3.2 Interviews

We chose to employ a qualitative approach based on interviews in this research phase, as it enabled us to gather insights that could not be obtained through the questionnaire alone. During this stage, our focus shifted from general perspectives to a deeper understanding of professionals' experiences with technical onboarding, aiming to uncover their feelings, thoughts, and attitudes regarding software visualization.

*3.2.1 Data Collection.*
Following the recommendations from the literature on conducting interviews, we adopted a semi-structured script to gather data from the participants [24]. The interview commenced with an initial set of questions to explore the interviewees' backgrounds in software development, alongside broader questions about the study topics, namely, technical onboarding and software visualization. The script was then adapted to incorporate specific questions tailored to the roles and experiences of the interviewees. For instance, project managers and leaders were presented with a particular set of questions, while software professionals in different roles were provided with a distinct set of questions. The final script comprised 55 adaptable questions, concentrating on experiences, insights, and practices related to technical onboarding and the use of software visualization techniques and tools in diverse software project scenarios. The main questions of the final script are exemplified by Table 2.

Following the questionnaire data collection, 14 participants expressed their willingness to participate in the qualitative phase of the study by providing their contact emails. Out of these, interviews were successfully conducted with 7 participants who confirmed their availability and consented to have their interviews recorded and transcribed. The interviews were conducted using the Google Meet platform in the latter half of March 2023. The interview duration varied, ranging from 15 to 35 minutes, with each participant offering different levels of detail and insights about their experience. The cumulative duration of the recorded interviews was 2 hours and 57 minutes. The interviews were then transcribed, and the text was submitted to data analysis.

*3.2.2 Data Analysis.*
We employed thematic analysis, following well-established guidelines [9] to delve into the extensive narratives provided by interviewees in response to our questions. This method enabled us to extract and identify recurring themes and patterns within the qualitative data, offering a comprehensive understanding of participants' experiences with technical onboarding and data visualization.

**Table 1: Survey Questionnaire**

*About You and Your Workplace*

1. How long have you worked in the software development industry?

2. What is your title or role in the software development industry?

3. What is the legal status of the organization you work for (private, public, for-profit, non-profit)?

4. What is the industry sector of the organization you work for?

5. What is the size of the organization where you work?

*Onboarding*

6. Does your team, department or organization have defined onboarding processes in general?

7. Does your team, department or organization have defined technical onboarding processes?

8. Are these processes, if present, documented and accessible to you and your team?

9. Which options best define the onboarding processes adopted by your team, department or organization?

10. If you have gone through a technical onboarding process in your team, department or organization, how satisfactory would you rate the experience?

*Software Visualization*

11. Did you know the term "software visualization" as described at the beginning of this page before completing this questionnaire?

12. What types of software visualization tools have you come across?

13. Does your team, department or organization use or have used software visualization tools as part of your onboarding processes?

14. If your team, department or organization uses or has already used software visualization tools in your onboarding processes, how would you rate your experience with it?

15. In your team, department or organization, were you responsible or involved in the decision to apply software visualization tools?

16. In what situations did you directly use software visualization tools?

17. If you have used software visualization tools directly, how satisfactory was your experience?

**Table 2: Interview Script Summary**

1. What is your current job title and role in your team?

2. What activities do you perform in this role?

3. How long have you been in this role?

4. Have you been introduced to the current team for this same role? If not, what would it be back then?

5. Are you now, or have you ever been in a leadership role in software teams?

6. Are you responsible or involved in your team's onboarding process?

7. Did you participate in defining the process or its application?

8. What tools are used in your team's current onboarding process?

9. Is this process already established, under construction, or constantly evolving?

10. Is there any evaluation or feedback mechanism for the onboarding process?

12. Are you satisfied with your team's current onboarding process? Why?

12. What are the most critical aspects of this process?

13. Have employees been slow to show themselves productive in this team?

14. Describe the best onboarding process you've participated in. What was the best aspect of it? What tools were used?

15. Did it take you a while to feel productive on that team?

16. Are you familiar with the concept of Software Visualization?

17. Have you ever used any Software Visualization tools? At what moment in the project? Did it satisfy the requirements of your task at the time?

18. What features were taken into consideration to include the tools in this process?

19. What features a tool would need in order to be adopted in your team's onboarding process?

20. What other tools would it need to be integrated to, if any?

21. Would this tool be used collaborative with other people? In what ways?

22. Would the licensing be an important factor for the tool to be adopted by your team? Would you prefer commercial or open-source tools, and why?

In this process, our focus was on examining topics closely related to the interviewees' experiences and the impact of onboarding on their team integration. We also delved into the techniques and software visualization tools used in this context, as well as the quality of the experience resulting from their use.

Furthermore, we conducted a comprehensive analysis by cross-referencing the data collected from both the questionnaires and interviews. This integrative approach allowed us to address the research questions effectively and formulate recommendations concerning the utilization of software visualization tools in the technical onboarding process.

### 3.3 Ethics

In accordance with ethical principles, participants were presented with a comprehensive explanation of the research's scientific objectives and the intended use of their data for research purposes before they answered the questionnaire. All participants provided their voluntary consent to partake in the study. Throughout the survey phase, our research did not collect any personal information from participants, such as names, email addresses, or employer details, to ensure their anonymity. The only exception was the contact email voluntarily provided by participants who were willing to participate in the subsequent stage of the research. For the interview phase, participants who had previously agreed to take part were once again reminded of the research's scientific objectives, and their identities, as well as any references to their employers, were anonymized.

## 4 FINDINGS

In this section, we present the main findings obtained from our analysis and provide the evidence used to answer our specific research questions. We integrated the insights gathered from both the questionnaires and the interviews to enhance the overall comprehension of the findings. It is worth noting that certain quotations may read oddly as they were translated to English.

### 4.1 Demographics

We obtained data from a sample of 41 software professionals working in a variety of contexts within the software industry. The initial groups of questions in the questionnaire provided us with insights into the respondents' profiles in terms of their roles and experience in the software development field. These profiles exhibited significant diversity, as illustrated in Figure 3.

The predominant role within our sample was software developers, specifically software programmers. They constituted a group of 14 individuals, making up 34% of the sample. Furthermore, our sample encompassed a range of other roles, such as designers, software architects, and software testers. In terms of software professionals in leadership positions, ten individuals (24%) held roles as software project managers or technical team leaders. In terms of company size, the majority of respondents (95%) indicated that they work in large companies with over 1000 employees, while 2 participants reported being employed by organizations employing between 500 and 999 professionals.

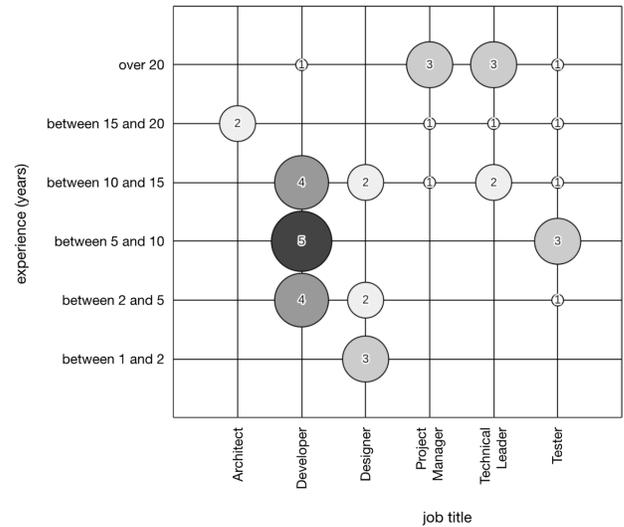

**Figure 3: Participant's Profile**

### 4.2 Onboarding Process in the Software Industry

When exploring onboarding processes within our sample, 58% of participants (24 out of 41) indicated that their company as a whole has established onboarding processes that are to be followed across the entire organization. These onboarding processes appear to be more generic in nature, typically commencing and concluding within the human resources domain. Additionally, 37% (15 out of 41) reported that their teams, in particular, have crafted onboarding processes that encompass training tailored to the unique aspects of the software being developed. Furthermore, the majority of the sample (69%) indicated that they had experienced well-documented and easily accessible onboarding processes, either at the company or team levels. Only two respondents in the sample mentioned that their teams lack clearly defined onboarding processes, and the integration of new professionals into software projects occurs in an ad-hoc manner, with minimal to no planning.

Additionally, we explored the experience of our sample with tools employed during the onboarding process, and the most commonly used tools reported by our participants were static documents and tutorials (34 responses), synchronous group presentations, either in-person or remote (25 responses, and synchronous one-on-one presentations, conducted in person or remotely (21 responses). In total, respondents indicated a total of 9 different tools and techniques, as illustrated in Figure 4.

In terms of their onboarding experience, the majority of respondents expressed satisfaction with their integration into their teams. A total of 18 professionals reported being satisfied, and five were very satisfied, with only one respondent expressing dissatisfaction. The main reasons for these responses were their ability to understand the source code and feeling productive after some time in the project.

The qualitative insights obtained from interviews allowed us to gain a more nuanced understanding of the sentiments and attitudes of software professionals regarding their onboarding experiences. In general, participants expressed diverse views about their satisfaction levels with onboarding processes and highlighted the challenges they face and potential areas for improvement.

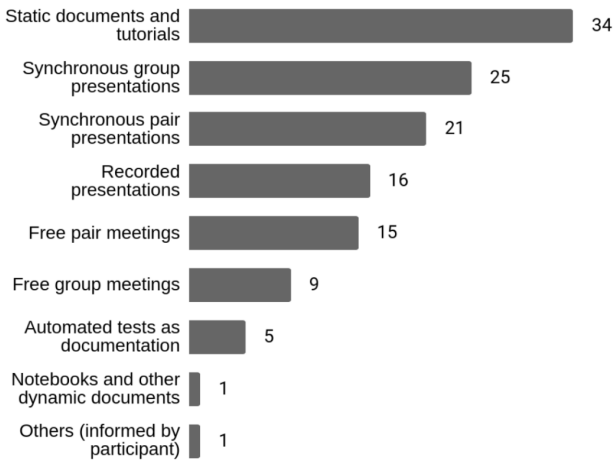

Figure 4: Tools Commonly Used in the Onboarding Process

For software professionals in leadership positions, particularly those responsible for designing and executing the onboarding process, a major challenge pertains to the significant time and workforce required for knowledge transfer, which can result in extended onboarding periods. This extended onboarding process may impact the time it takes for newcomers to reach full productivity in their activities. The quotations below were extracted from interviews and refer to this finding:

> "What I wasn't satisfied with was the time it took, you know? The onboarding process became overly extended, and I think it was too time-consuming." [P2]

> "This duration is still too long. It costs money, time, and effort because it's not very efficient." [P3]

> "Some people take a bit longer to start contributing (to the work), but the average is around two sprints, so about a month." [P5]

> "The thing is, we have to frequently allocate some team members to provide more support to the new team member, and it incurs delays." [P6]

As from the perspective of newcomers who are not in leadership roles and experience the onboarding process as users rather than planners, the onboarding process is beneficial in terms of team integration and facilitates the understanding of the software project's intricacies. However, these participants have observed that the use of documents and other static methods often leads to outdated information, causing long-term disadvantages, as illustrated in the quotations below.

> "The system is from a very complex context. The sessions I took with the designer have been quite helpful. It would be much more difficult to understand the system on my own just by reading the code." [P7]

> "If I were to onboard you, I would use the documents we have, but I would also mention that much of what is in there is outdated. I believe we no longer use a lot of it. Perhaps there is no one responsible for keeping that documentation up to date. I think there should be a more personal guidance. If there were a 're-onboarding' to review things again, it would be interesting." [P4]

### 4.3 Software Visualization

The concept of software visualization is not widely recognized within the industry. Before responding to our questionnaire, more than 68% of the respondents (28 out of 41) were not familiar with the term "software visualization." However, after being provided with a definition of the term, participants were able to identify various software visualization techniques and tools they had used. Among the most commonly mentioned ones were graphs and conceptual maps (19 responses), automatically generated UI flows (18 responses), and dependency graphs (17 responses), as shown in Figure 7.

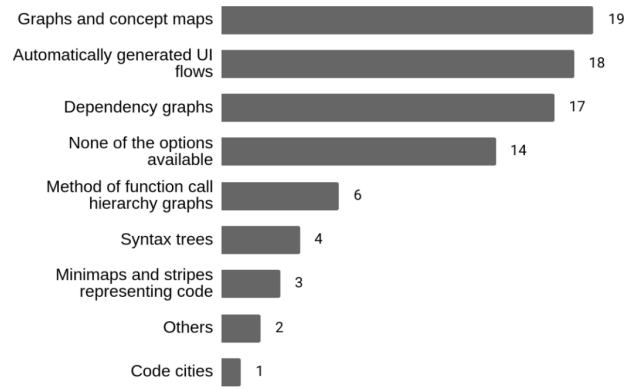

Figure 5: Software Visualization Known by Software Professionals

In terms of using these tools during onboarding, 54% of the respondents mentioned that they do not use them, 39% were unsure, and only 7% (3 respondents) confirmed they have experienced such tools and rated their experience as satisfactory. Among these, two respondents stated that they successfully employed the tools for onboarding purposes and found them to be effective.

Throughout the interviews, participants shared more in-depth insights into the relevance of software visualization in the software industry. In general, they revealed the application of such tools, with some applying them exclusively during the onboarding process, while others integrated them throughout the entire development lifecycle. Furthermore, some interviewees emphasized the importance of software visualization for program comprehension in the context of legacy code.

> *"I had used tools for creating diagrams and code repository visualization tools. I used these tools for generating both UML and dependency graphs."* **[P2]**

> *"We used it not only in onboarding but also in code assessment. (...) The architects would design and then reverse-engineer the code to check for compliance with the auditing process."* **[P3]**

When it comes to desired functionalities in software visualization tools, participants emphasized the importance of integrating these tools with activity management systems to streamline software project management processes. In addition, software professionals expressed the need for greater flexibility in these tools, including more dynamic features. They desire the capability to remove irrelevant elements within visualizations and customize the layout of code elements, which would help in keeping the onboarding process up to date.

> *"The first aspect would be user-friendliness since onboarding is a one-time event. In my experience, such tools are effective for providing an overview or for aiding in specific tasks, but they can sometimes create overly complex diagrams."* **[P2]**

> *"The tools I utilized had a somewhat unconventional code reversal process for documentation, which often necessitated manual intervention."* **[P3]**

Our interviewees emphasized another crucial aspect of software visualization tools: the potential for collaborative use among users, both in real-time and asynchronously. The collaborative use of these tools, especially for maintaining knowledge about the software project, enhances opportunities for integration among professionals. For instance, tools like Jira, which connect professionals from various roles, facilitate this collaborative work according to the participants.

> *"then you have integration for both reverse code engineering, with the IDE, test case tools, and so on. These integrations make the software development lifecycle process much smoother."* **[P3]**

> *"Integration with Jira would be quite interesting. (...) Having a team-level tool with different people looking at it would be highly beneficial."* **[P7]**

In general, we identified that regardless of the tools employed for program comprehension, there is a prevalent concern among managers and leaders about enhancing the productivity of software professionals, particularly in terms of the time it takes for them to reach peak performance after joining the team. This emphasis on productivity remains a common thread in discussions, regardless of the specific context of onboarding processes or software visualization tools during the software development life cycle, as demonstrated in the quotations below.

> *"It is estimated that software professionals currently spend up to 220 hours on onboarding. (...) When we consider the time it takes for professionals to become productive, there is a perception of delayed productivity. During retrospective meetings, it becomes evident that the onboarding process, especially within the first two sprints, requires improvement."* **[P3]**

> *"Implementing a more formalized feedback process from those who have undergone onboarding could contribute to greater productivity before diving into work."* **[P5]**

## 5 DISCUSSION

Building upon our data analysis, we aim to provide recommendations for leveraging software visualization tools to enhance the effectiveness of technical onboarding sessions and the creation of strategies to support software development teams. Hence, in this section, we primarily address the research questions posed in this study and subsequently delve into the implications of our findings.

### 5.1 How do software professionals navigate technical onboarding and the utilization of software visualization tools within the software industry?

To tackle our main research question, we delved into the details of our specific research questions, with each of them focusing on distinct aspects of our comprehensive investigation. Additionally, we integrated relevant findings from the existing literature and discussed how our results correlate with prior research on technical onboarding and software visualization.

*5.1.1 How familiar are software professionals with software visualization tools?*
Overall, the level of familiarity of software professionals with software visualization tools is relatively low. In many instances, software professionals are not aware of the concept of software visualization or the specific tools available in this domain. This evidence aligns with previous discussions in the literature on program comprehension [34]. Despite some participants showing limited familiarity with specific visualization tools mentioned in the study, it is evident that the concept of software visualization itself is not widely recognized among industry professionals.

Our findings also align with existing literature in considering a disconnect between the concept of software visualization and the practical tools and techniques commonly used in the software industry [21, 22]. Despite a small number of participants mentioning prior experience with software visualization tools, the overall lack of awareness regarding their broader applicability and the benefits highlighted in academic research creates a significant obstacle for professionals when contemplating their integration into their work processes. Our results indicate that when practitioners do use these tools, it is typically for specific and isolated tasks, often without a comprehensive understanding of the additional advantages or established best practices in the field [22, 34, 40].

*5.1.2 How frequently do software companies incorporate software visualization tools into their technical onboarding processes?*
The frequency with which software companies incorporate software visualization tools into their technical onboarding processes varies among the respondents in our study. Some companies have integrated these tools as a routine part of their onboarding process, while others do not use them at all. Confirming results discussed in the literature [22], the adoption of software visualization tools in the industry remains limited.

Our results emphasize a notable gap in the use of such tools for system documentation and the facilitation of technical onboarding sessions [34]. This scenario underscores the potential of strategies focused on systematizing and formalizing onboarding processes, as well as knowledge management, to incorporate software visualization techniques to effectively promote and facilitate knowledge transfer within software teams [4].

*5.1.3 How do software professionals assess the effectiveness of software visualization tools in improving their comprehension of source code during the onboarding process?*
Software professionals had varying opinions about the effectiveness of software visualization tools in their onboarding processes, with some highlighting the benefits while others pointed out potential challenges. Some participants expressed satisfaction and found these tools helpful in understanding the code and becoming productive more quickly. However, others raised concerns about the time and effort required to use these tools effectively.

Our findings indicate that visualizations could sometimes become overly complex, and manual adjustments were often needed to make them more useful. Furthermore, there were concerns about static tools becoming outdated over time, which could result in long-term disadvantages. In this context, we observed a need for usability enhancement in such tools [28], particularly because the onboarding process is expected to be simple and efficient, minimizing the workload and assisting professionals in achieving satisfactory productivity sooner rather than later.

## 5.2 Recommendations

Drawing upon the interpretation of our results and a comparison with the findings in the literature, we can formulate the following recommendations for practitioners:

- *Systematization of Onboarding with Differentiation Between Institutional and Technical Onboarding*: As discussed in this study, the onboarding process can either follow established models or lack them altogether. It is worth noting that two distinct types of onboarding are applicable to the software industry: events and training for institutional integration and technical onboarding within the software development industry team. While most companies have well-defined and documented processes for the first type, many do not formalize processes for the second type. This can lead to variations in productivity and professional dissatisfaction among newcomers. We recommend establishing well-documented processes for technical onboarding, taking into consideration the project lifecycle, workflow, as well as architectural aspects, and program comprehension.
- *Capacitating Software Professionals in using Software Visualization Tools*: To ensure the effective application of software visualization tools in the onboarding process, it is essential to enhance professionals' knowledge of the techniques and tools. This can be achieved through training activities. Regular updates and the continuous use of training programs are encouraged. Additionally, to facilitate the process of technical onboarding and documentation, instructions for using the tools adopted in each project can be incorporated into README files within their source code repositories.
- *Introduction of Visualization Tools in the Software Project Lifecycle*: Our findings encourage the adoption of software visualization tools not only to support technical onboarding but also to enhance program comprehension throughout the software project life cycle. For ongoing projects, these tools can be invaluable in elucidating system architecture, tracking software evolution, demonstrating the relationship between code variability and planned activities, and serving as interactive documentation for new team members. Additionally, new software projects can harness the power of software visualization to assist managers and clients in grasping the cost of requirements.
- *Selecting and Evaluating Software Visualization Tools*: Just as with any supporting resource, it is advisable to assess the software visualization tools to be incorporated into the software project using empirical methods. The selection process should take into account the project state, the specific aspects of the systems, the familiarity of the team with the visualization tools, the availability of integration with other tools, and the performance and robustness of the tools when applied in contexts similar to those encountered in the project.

## 5.3 Limitations and Threats to Validity

In regard to our survey, it is essential to acknowledge that our sample size is limited, which restricts the extent to which our findings can be generalized. Additionally, our research is subject to inherent methodological limitations of qualitative research since our results depend on our interpretation of the data collected during interviews. To address this potential threat to validity, we applied a triangulation approach by combining data from the questionnaires and the interviews. Furthermore, we heavily relied on participant quotations to consistently conduct our thematic analysis.

Despite these limitations, we believe researchers and practitioners can derive valuable insights from our findings and apply the knowledge gained in their specific contexts. This transfer of knowledge is similar to the process employed when exploring and applying findings from case studies, grounded theory, and other methods that do not permit generalizations.

Therefore, we understand that our findings can stimulate discussions and provide valuable insights to practitioners when considering the advantages of technical onboarding and the effective use of software visualization tools in software projects.

### 5.4 Future Work

Our research has unveiled several promising paths for future investigations within the domain of technical onboarding in the software industry and the utilization of software visualization tools. Some key research directions we are considering for future exploration include:

- Assessing how different software visualization tools align with the specific needs of software professionals during the onboarding process. This evaluation would take into account various contexts, such as agile methodologies, remote work, hybrid work setups, and global software development.
- Comparing the utilization of these tools in the software industry during onboarding stages and in the day-to-day project activities.
- Exploring the productivity and satisfaction levels among newcomers considering those who incorporate software visualization tools during onboarding and those who integrate them into their daily work routines.

These potential research directions hold the promise of advancing our understanding of the interplay between technical onboarding and software visualization, offering valuable insights for both the academic and industrial sectors.

## 6 CONCLUSION

In this study, our primary goal was to investigate the intersection of technical onboarding in the software industry and the application of software visualization tools. Our objective was to provide valuable insights that can assist software practitioners in refining their onboarding processes.

Our research provides important findings about the state of technical onboarding in the software industry and the limited awareness of software professionals regarding software visualization tools. We identified the need for systematic onboarding processes, differentiation between institutional and technical onboarding, and formalized training to enable software professionals to use software visualization tools effectively. These insights pave the way for enhancing the onboarding experience and deepening program comprehension.

Additionally, our findings emphasize the significance of software visualization in facilitating the understanding of complex software projects, not only during onboarding but throughout the entire project lifecycle. In conclusion, this study is a foundational piece of evidence for further research and discussion in technical onboarding, software visualization, and knowledge transfer within software development teams.